\documentclass[twocolumn,printnumbers,amsmath,amssymb,showpacs]{revtex4}

\newcommand{\be}{\begin{equation}}
\newcommand{\ee}{\end{equation}}
\newcommand{\ba}{\begin{eqnarray}}
\newcommand{\ea}{\end{eqnarray}}
\usepackage{graphicx}
\usepackage{dcolumn}
\usepackage{bm}

\begin{document}

\title{Attenuation of shear sound waves in jammed solids}


\author{Vincenzo Vitelli} 
\affiliation{Instituut-Lorentz, Universiteit Leiden, Postbus 9506, 2300 RA Leiden, The Netherlands.}

\begin{abstract}

We study the attenuation of long-wavelength shear sound waves propagating through model jammed packings of frictionless soft spheres interacting with repulsive springs. 
The elastic attenuation coefficient, $\alpha(\omega)$, of transverse phonons of low frequency, $\omega$, exhibits power law scaling as the packing fraction $\phi$ 
is lowered towards $\phi_c$, the critical packing fraction below which rigidity is lost. The elastic attenuation coefficient is inversely proportional to the scattering mean free path and 
follows Rayleigh law with $\alpha(\omega)\sim \omega^4 (\phi - \phi_c)^{-5/2}$ for $\omega$ much less than $\omega^* \sim (\phi - \phi_c)^{1/2}$, the characteristic frequency scale above which the energy diffusivity and density of states plateau. This scaling of the attenuation coefficient, consistent with numerics, is obtained by assuming that a jammed packing can be viewed as a mosaic composed of domains whose characteristic size $\ell^ * \sim (\phi-\phi_c) ^{-1/2}$ diverges at the transition. 

\end{abstract}

\maketitle

\section{\label{sec:intro}Introduction}

The challenge of attaining a robust understanding of the elasticity and vibrational dynamics of amorphous solids continues to be an elusive goal in material science \cite{Elliott_book}. 
One of the main difficulties arises from the uncertainty surrounding the application of the time honored formalism of continuum mechanics to disordered structures characterized by 
strong spatial inhomogeneities and by a non affine response to external perturbations, which often acts as a precursor to elastic instabilities \cite{Leonforte2,matthieu2,Gurevich}. The limitations
of our current understanding are manifest when one considers the slow pace of progress in achieving a satisfying picture of the low temperature properties of glasses 
(such as the ubiquitous presence of a plateau in their thermal conductivity) or of sound propagation in disordered solids such as emulsions, foams and granular media \cite{Phil81,Jia,Liu,Lub03,Sheng91,Allen93,Pohl87,XLiu,sch07,Parshin}. 

Several authors have been intrigued by the prospect of a unified description of these apparently diverse phenomena that relies on 
general principles common to molecular glasses as well as amorphous packings of bubbles or grains \cite{O'Hern03,Leonforte2,matthieu}.
The generic starting point, often invoked in the context of simple models and computer simulations, is the existence of a characteristic {\it mesoscopic} length scale,
intermediate between the microscopic size of the basic components and the sample size, below which continuum elastic theories
cease to be valid \cite{Leonforte2,matthieu,matthieuleo}. A consequence of the breakdown of classical elasticity is that the vibrational degrees of freedom characteristic of amorphous solids cannot always be identified as simple plane waves. Thus, mechanical and transport properties controlled by the vibrational dynamics of amorphous solids differ markedly from their crystalline counterparts.
   
Experimental progress in measuring elastic wave scattering has provided reliable nondestructive methods to probe the structure of amorphous materials at different length scales \cite{page}. 
By varying the frequency or the wavelength of the incident wave, different spectral regimes of the vibrational dynamics of the disordered samples become experimentally accessible. 
This provides an identikit of the vibrational degrees of freedom of the amorphous solids under investigation and paves the way to understanding how they emerge from 
characteristic length scales inherent to these disordered structures. 

In this work we concentrate on how sound propagates through jammed packings of frictionless soft repulsive spheres whose density can be varied above the critical threshold necessary to insure mechanical rigidity \cite{O'Hern03}. 
This simple and concrete model system can be readily generated on a computer and captures several of the generic properties of amorphous materials, such as the onset
of excess vibrational modes at a characteristic frequency $\omega^*$ \cite{Leo1}, above which the scattering mean free path and the wavelength are comparable \cite{Xu09,Vitelli10}. The characteristic frequency $\omega^*$ heralds a spectral regime of strong scattering characterized by a small and frequency independent energy diffusivity. This feature is also seen in experiments carried 
out on glasses as well as porous media  \cite{page,Graebner}. 

A unique and intriguing feature that motivates our choice of jammed solids over other model systems is the fact that their elastic properties exhibit scaling behavior with packing fraction or connectivity above the critical 
density $\phi=\phi_c$ (Point J) at which the particles are just touching each other ~\cite{O'Hern03}. In particular, the mesoscopic length $\ell^ * \sim (\phi-\phi_c) ^{-1/2}$ above which 
continuum elasticity is expected to hold in this model, can be made to diverge as the unjamming transition is approached by progressively decompressing the sample towards point J \cite{matthieu,matthieuleo}.
However, some uncertainty still exists regarding the existence and nature of this diverging length scale.

A recent effective-medium investigation of phononic transport in random networks of harmonic springs captured several properties shared by jammed packings of soft spheres \cite{Mattnew}. The random network model predicts a sharp crossover at $\omega^{*}$ in the energy diffusivity from a plateau to a low-$\omega$ Rayleigh scattering regime, with a pre-factor that scales with connectivity and eventually vanishes at the isostatic point. Despite its success, this mean field approach does not yield any transport signatures of the characteristic length scale $\ell ^{*}$ that controls the fluctuations in the microscopic structure. 

The diffusivity plateau above $\omega^*$ was also studied directly for the jammed packings by evaluating the Kubo formula without recourse to a mean field approximation \cite{Vitelli10,Xu09}. However, no attempt was made in Ref. \cite{Vitelli10,Xu09} to extract the packing fraction dependence of Rayleigh law for $\omega\ll \omega^{*}$.  

In this article, the scaling of the pre-factor to Rayleigh law is determined by studying the attenuation coefficient of long-wavelength transverse acoustic phonons which carries a vestige 
of the important length scale $\ell^{*}$ associated with the unjamming transition. The numerical data of Ref. \cite{Vitelli10,Xu09} is re-analyzed and shown to be consistent with the present analysis,
but not with mean field calculations because the latter do not yield the length scale $\ell ^{*}$. 
Our scaling analysis of the attenuation coefficient provides insights on the structure of the jammed solids. In particular, it 
suggests a simple view of the jammed solid as a mosaic composed of domains whose characteristic size is of order $\ell ^{*}$. 

In addition to its relevance as a probe of the microstructure of a jammed solid, the present analysis suggests a methodology to infer the fabric of a granular media from experimental investigations of sound propagation under external loading \cite{Jia,Liu}. 
 
\section{\label{sec:intro} Review of the elasticity of jammed sphere packings}

The model jammed solids studied in this work are amorphous packings of frictionless spheres generated by conjugate-gradient energy minimization using the protocol described in Ref. \cite{O'Hern03}.
The harmonic pair potential $V(r_{ij})$ between particles is purely repulsive: 
\begin{eqnarray}
V(r_{ij})&=&\frac{\epsilon}{2}(1-r_{ij}/\sigma_{ij})^{2}  \quad \quad \!\!  \text{if} \quad r_{ij}<\sigma_{ij} \nonumber \\
V(r_{ij})&=& 0  \quad \quad \quad \quad \quad \quad \quad \quad \text{if}  \quad r_{ij}>\sigma_{ij} \quad \!\!\!  ,
\label{eq:C}
\end{eqnarray}
where $r_{ij}$ denotes the distance between the centers of particles $i$ and $j$ and
$\sigma_{ij}$ is the sum of their radii.

At the unjamming transition $\phi = \phi_c$, the average coordination number, $z$, increases discontinuously from zero to the ``isostatic" value $z_c=2D$, where $D$ is the dimensionality of the sample ~\cite{alexander}.  Above $\phi_c$ at $\Delta \phi \equiv \phi-\phi_c$, the average coordination number exhibits power law scaling ~\cite{Durian95,O'Hern03}:
\begin{eqnarray}
\Delta z \equiv z-z_c \sim \Delta \phi^{1/2} \!\!\!\!\! \quad . \label{eq:scaling00} 
\end{eqnarray}   
In addition, the shear modulus scales with $\Delta \phi$ according to \cite{Durian95,O'Hern03}:
\begin{eqnarray}
&G& \sim \Delta \phi^{1/2}  \!\!\!\!\! \quad , \label{G}
\end{eqnarray}  
whereas the bulk modulus $B$ is only weakly dependent on packing fraction near the jamming point. The anomalous scaling of the elastic moduli reflects the importance of non-affine deformations \cite{wouter}.

The transverse and longitudinal speeds of sound $v_{t}$ and $v_{l}$ are proportional to the square root of the shear and bulk moduli, which implies that only the transverse speed of sound scales with $\Delta \phi$ according to
\begin{eqnarray}
&v_t& \sim \Delta \phi^{1/4}  \!\!\!\!\! \quad .  \label{eq:scaling2bis}  
\end{eqnarray}  
Upon substituting Eq. (\ref{eq:scaling2bis}) into the Deybe formula, the ratio of the transverse to the longitudinal density of states at low frequency diverges as $\Delta \phi  \rightarrow 0$. Above a characteristic frequency 
$\omega^{*} \sim {\Delta \phi}^{1/2}$ the density of states exhibits a plateau that extends all the way down to zero frequency at point J \cite{Leo1}. That is to say, in a jammed solid right at the threshold of rigidity 
the familiar Deybe regime is completely absent and the acoustic waves, which are the familiar excitations of continuum elastic media, are suppressed. 

An explanation of this puzzling phenomena was proposed by Wyart and collaborators who have suggested that a characteristic length scale $\ell ^{*}$, which diverges as $\phi$ approaches $\phi_c$, 
controls the elasticity of jammed solids \cite{matthieu,matthieuleo}. Here, we briefly review their ``cutting argument" originally proposed to obtain the packing fraction dependence of $\ell ^{*}$. 

Consider the mechanical stability   
of a jammed packing at a packing fraction $\phi > \phi_c$, such that there is a density of inter-particle contacts $\Delta z$ in excess of the minimum isostatic value $z_c$ necessary to ensure stability. 
The properties of a  continuum elastic medium should remain invariant upon cutting it in two parts. If you cut a spherical blob of radius $\ell$ out of a jammed medium, a number of contacts equal to the area $\ell^{d-1}$
of the blob will be broken. However, the excess number of contacts that exist in the blob is given by  $\ell^d \!\!\!\!\! \quad \Delta z$. One obtains the size $\ell^{*}$ of the smallest blob that will remain rigid upon cutting, by 
balancing the number of broken contacts with the excess ones. This leads to  \cite{matthieu,matthieuleo}
\begin{eqnarray}
\ell^{*} \sim \frac{1}{\Delta z} \sim \frac{1}{\Delta \phi^{1/2}}  \!\!\!\!\! \quad ,  \label{lstar}  
\end{eqnarray}  
where Eq. (\ref{eq:scaling00}) was used to obtain the explicit dependence on packing fraction. According to this elegant argument, below the length scale $\ell ^{*}$ the medium is to be considered floppy and hence unable to support
the acoustic (quasi) plane waves that give rise to the Deybe counting in the vibrational density of states. Long wavelength phonons with $\lambda \gg \ell ^{*}$ can still be supported because on those length scales the jammed packing is rigid and can be described as a continuum elastic medium. 

The conclusions of this simple argument are consistent with the numerical studies carried out by Ellenbroek and collaborators who probed the fluctuations in the force response of a jammed solid to inflating the diameter of a single particle at the center of the sample \cite{wouter}. The spatial range of these fluctuations is given by $\ell^{*}$ and grows upon approaching the jamming point.   

Despite this progress, there is still uncertainty regarding the exact character and the very existence of this important length scale. In particular it is not clear whether $\ell^{*}$ can be viewed as a structural signature of the unjamming transition or merely as a dynamic length scale that appears in the vibrational dynamics of these fragile solids. 

In the next section, we will address this issue by studying the packing fraction dependence of the scattering attenuation coefficient of long wavelength phonons. Such waves can be excited by introducing in the jammed system an oscillatory perturbation of very low frequency to probe dynamically the divergent fluctuations observed in Ref. \cite{wouter}. 

\section{\label{sec:intro} Packing fraction dependence of the attenuation coefficient}

\subsection{\label{sec:scaling} Scaling analysis.}

Consider a shear sound wave of very low frequency $\omega$ propagating through a jammed packing with a well defined transverse speed of sound,  $v_t$. 
In this section, we develop simple physical insights that allow us to infer the presence of the diverging
length scale $\ell^*$ from the elastic scattering of these long wavelength shear waves.  

The approach pursued here is different from previous studies which have sought signatures of this length scale in the vibrational dynamics around and above $\omega^{*}$.
The regime $\omega \ll \omega^*$ has generally been considered uninteresting because the density of states of weakly scattered phonons $n(\omega)$ is obtained
by simply counting how many waves you can fit in a three 
dimensional box. As point J is approached, the shear modulus becomes much softer than the bulk modulus (see Eq. (\ref{G}) and the following comment) and the spectrum is dominated by shear waves. The transverse density of states $n_t(\omega)$ takes up the generic form 
\begin{equation}
n_t(\omega)= \frac{\omega^2}{{v_t}^{3}} 
\label{eq:def0}
\end{equation}
The packing fraction dependence of the pre-factor of Deybe law is entirely determined by the scaling of the shear modulus in Eq. (\ref{G}) and reads
\begin{equation}
n_t(\omega)= \frac{\omega^2}{{\Delta \phi}^{3/4}} 
\label{eq:def00}
\end{equation} 

Equation (\ref{eq:def0}) is merely what you expect for a perfectly ordered solid, hence it does not carry 
distinctive signatures of the amorphous structure of a jammed packing. This has prompted various investigators to seek indirect evidence 
for $\ell^*$ at higher $\omega$ where the density of states departs dramatically 
from Deybe behavior and exhibits a characteristic plateau. 

Here we show that the
sound attenuation coefficient $\alpha(\omega)$ is a more sensitive spectral measure than the density of states and can display anomalies as a result of 
elastic scattering from structural inhomogeneities even in the low $\omega$ regime for which the density of states approximately follows Deybe law. 

In order to define the sound attenuation coefficient, denote by $I_0$ the initial intensity (proportional to the elastic energy density) and by $I(x)$ the reduced intensity attained by 
the wave after traveling a distance $x$ through
the sample. The coefficient of sound attenuation $\alpha(\omega)$ can then be extracted from the relation
\begin{equation}
I(x) = I_{0} e^{-\alpha(\omega) x} 
\label{eq:def1}
\end{equation}
In what follows we focus our attention to the simpler case in which sound attenuation is primarily triggered by elastic  scattering. This restriction precludes us from considering the effects of absorption and dissipation due to friction which can be significant for a real granular packing \cite{Jia2,Jia3}.   

The elastic scattering contribution to the attenuation coefficient is inversely proportional to the scattering mean free path $\ell_{s}(\omega)$  \cite{west}
\begin{eqnarray}
\alpha(\omega)\sim \frac{1}{ {\ell_{s} }(\omega)}
\label{eq:def2}
\end{eqnarray}
Sound attenuation in granular media was observed to be dominated by transverse waves \cite{Jia1,Jia2}. 
According to Rayleigh's law, the mean free path $\ell_{s} (\omega)$ of a tranverse sound wave is given by the generic form \cite{west,John}
\begin{eqnarray}
\ell_s (\omega)\sim \left(\frac{v_t}{ \omega}\right)^4 \!\!\!\! \quad \left( \frac{1}{D} \right)^3
\label{eq:def3}
\end{eqnarray}
where $D$ is the length scale that characterizes the disordered structures responsible
for the scattering. As anticipated, the presence in Eq. (\ref{eq:def3}) of the multiplicative factor $D^{-3}$, which does not appear in Eq. (\ref{eq:def0}), makes the mean free path $\ell_s (\omega)$ 
a more sensitive probe of amorphous structures than the density of states $n_t(\omega)$.

In polycrystalline solids, the length $D$ is given by the typical distance between grain boundaries. If the characteristic length scale of structural inhomogeneities in a jammed media had no 
packing fraction dependence, Eq. (\ref{eq:def3}) would lead to $\ell_s (\omega) \sim \omega^{-4} \Delta \phi$. Instead, we make the working assumption that the divergent fluctuations of size $\ell^ * \sim (\phi-\phi_c) ^{-1/2}$,  observed in response to a static perturbation
in Ref. \cite{wouter}, are still crucial when the system is probed dynamically by sending low $\omega$ elastic waves through it. 

Upon setting the speed of sound equal to the square root of the shear modulus and $D \sim \ell^{*}$ in Eq. (\ref{eq:def3}), we obtain
\begin{eqnarray}
\ell_s (\omega) \sim  \frac{1}{\omega^4} \!\!\!\! \quad \frac{G^2}{{\ell^{*}}^{3}} 
\label{eq:def4}
\end{eqnarray} 
Substituting Equations (\ref{G}) and (\ref{lstar}) into Eq. (\ref{eq:def4}) gives the packing fraction dependence of the attenuation coefficient (or equivalently the inverse of the mean free path) 
\begin{eqnarray}
\frac{1}{\alpha (\omega)} \sim \ell_{s}(\omega) \sim \frac{1}{\omega^4} \!\!\!\! \quad {\Delta \phi} ^{\frac{5}{2}} 
\label{eq:dscale}
\end{eqnarray} 
Equation (\ref{eq:dscale}) suggests that in the weak scattering regime $\omega \ll \omega^{*}$, the absorption length $1/\alpha(\omega)$ vanishes as point J is approached in a way controlled by the anomalous scaling of the shear modulus and of the length scale $\ell ^{*}$, providing indirect evidence for the existence of the latter. Indeed, a simple physical picture of how elastic waves propagate through a jammed packing under loading is obtained from viewing this disordered solid as a mosaic of 
domains with characteristic size $\ell^ * \sim (\phi-\phi_c) ^{-1/2}$. The resulting mean free path $\ell_{s}$ is not simply given by the characteristic length $\ell ^{*}$. Instead, upon approaching point J, $\ell_{s}$ {\it decreases} as $\ell ^{*} \sim \Delta \phi ^{-1/2}$  {\it increases}, as described by Eq. (\ref{eq:def4}). This is consistent with the intuitive notion that the unjamming transition is accompanied by a gradual loss of rigidity that decreases the efficiency of the energy transport by plane wave excitations hence resulting in a vanishing pre-factor for Rayleigh law at point J.   

The noteworthy feature that distinguishes the present analysis is that it applies to frequencies well below $\omega^{*}$ where the scattering of transverse sound waves is weak,
whereas $\ell ^{*}$ was originally introduced to explain the plateau that exists in the density of states above $\omega^*$. Recent investigations indicate that the vibrational modes above $\omega^*$ are diffusive and the corresponding mean free path is controlled by a distinct length scale $\ell_d \sim \Delta \phi^{-\frac{1}{4}}$ (see Sec. \ref{test2}).
  
\section{\label{sec:packings} Numerical tests}  

In the previous section, we obtained the packing fraction dependence of the (inverse) attenuation coefficient in the Rayleigh scattering regime by introducing the characteristic length $\ell^{*}$ by hand. In order to lend some support to this phenomenological approach, we show that the final result in Eq. (\ref{eq:dscale}) is consistent with numerical evaluations of the Kubo formula for the energy diffusivity that do not rely on any {\it ad hoc} assumptions.

\subsection{\label{sec:packings} Energy diffusivity}  

In the weak scattering regime $\omega \ll \omega^{*}$, where Rayleigh's law holds, the energy diffusivity, $d(\omega)$, can be approximated as 
\begin{eqnarray}
d(\omega) \sim v_{t} \ell_{s} (\omega)
\label{eq:def6}
\end{eqnarray} 

Inspection of Eq. (\ref{eq:dscale}) and Eq. (\ref{eq:scaling2bis}) allows us to write
\begin{eqnarray}
d (\omega) \!\!\!\! \quad \omega^4 \sim {\Delta \phi} ^{\frac{11}{4}} \!\!\! \quad .
\label{eq:def7}
\end{eqnarray} 

In Ref. \cite{Vitelli10}, the energy diffusivity of an unstressed packing \footnote{The unstressed packings are obtained by replacing the interaction potential, $V(r_{ij})$, between each pair of overlapping particles with an unstretched spring with the same stiffness, $V^{\prime \prime}(r_{ij}^{eq})$, where $r_{ij}^{eq}$ is the equilibrium distance between particles $i$ and $j$. This corresponds to dropping terms depending on the first spatial derivative of the potential, $V^\prime$, in the dynamical matrix which increases the mode energy and frequency, because $V^\prime$ is negative for repulsive interactions.}, comprised of a 50/50 bidisperse mixture of $2000$ frictionless spheres with a diameter ratio of 1.4, was computed numerically using the Kubo formalism \cite{Allen93} 
\begin{equation}
d(\omega_{i})\equiv \frac{\pi}{3} \sum_{j} (\hbar \omega_{i})^{-2} \!\!\!\!\! \quad |\vec{S}_{ij}|^2 \!\!\!\!\! \quad \delta(\omega_i - \omega_j) ,
\label{eq:kubo5}
\end{equation}   
where $\vec{S}_{ij}$ denotes the energy flux matrix elements $\vec{S}_{ij}$. They read \cite{Allen93}
\begin{equation}
\vec{S}_{ij}= \frac{(\omega_i + \omega_j)^2}{4 \!\!\!\!\! \quad \omega_i  \!\!\!\!\! \quad \omega_j} \sum_{m n,\alpha \beta} (\vec{r}_{m}-\vec{r}_{n}) e_{i}(m; \alpha) \!\!\!\! \quad H_{\alpha\beta}^{mn} \!\!\!\! \quad e_{j}(n; \beta) \!\!\!\! \quad .
\label{eq:Sij}
\end{equation}
where $\vec{r}_{m}$ indicate the particle positions, $H_{\alpha\beta}^{mn}$ the dynamical matrix and $e_{i}(m; \alpha)$ the corresponding eigenvectors \cite{Ashcroft}. 

The numerical evaluation of Eq. (\ref{eq:kubo5}) carried out in Ref. \cite{Vitelli10} provided clear of evidence for a sharp energy transport crossover at a characteristic frequency $\omega^{*}\sim \Delta \phi ^{\frac{1}{2}}$ that separates the Rayleigh law regime from a diffusivity plateau. However,
the packing fraction dependence of the pre-factor of Rayleigh law was not investigated. We now proceed to re-analyze the numerical data obtained in Ref. \cite{Vitelli10}
to extract and test the critical exponent predicted by Eq. (\ref{eq:def7}) for the diffusivity in the low $\omega$ regime . 

\subsection{\label{results} Results}

The scaling analysis of the previous section was implicitly carried out in an infinitely large system for which vibrational modes are continuously distributed in the spectrum. However, in a cubic simulation box of size $L$ the low $\omega$ acoustic waves occur close to the discrete frequencies allowed by the linear dispersion : 
\begin{equation}
\omega_{i} = \frac{2 \pi v_t}{L} \sqrt{p^2+q^2+r^2}
\end{equation}    
where $\{p,q,r\}$ denote the Miller indexes and $v_t$ the speed of sound. This allows the calculation of the mode degeneracies predicted by classical elasticity \cite{Leonforte2}. 
For example the lowest lying mode $(\pm 1,0,0)$ should be 12-fold degenerate because there are two transverse directions for the six allowed wave vector of magnitude $2 \pi/L$. Similarly the  
second lowest lying mode would have a degeneracy number equal to $24$. In finite disordered samples these degeneracies are partially lifted even in the regime $\omega \ll \omega^{*}$ where 
the vibrational modes are indeed quasi plane waves. 

\begin{figure}
\includegraphics[width=0.5\textwidth]{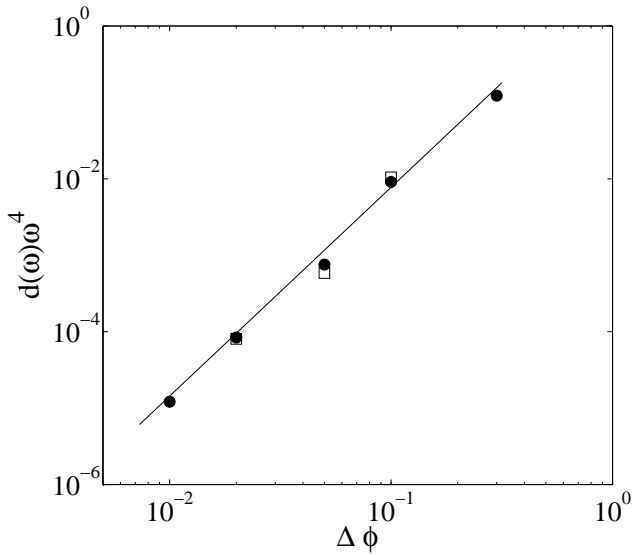}
\caption{\label{figsoft} Log-Log plot of 
$d (\omega) \!\!\!\!\! \quad \omega^4$ versus packing fraction $\Delta \phi$ for the lowest lying group of 12-fold degenerate modes at $\Delta \phi=0.3$ , $0.1$ , $0.05$ , $0.02$  and $0.01$ (circles) and for the second lowest lying group of 24-fold degenerate modes at $\Delta \phi=0.1$ , $0.05$ , $0.02$ (squares). The slope of the continuous line is $11/4$. The data is reproduced from Fig. 1 of Ref. \cite{Vitelli10}.}
\end{figure}

The correct procedure to bin the data, is to average the diffusivity and frequency of each degenerate group of long wavelength phonons separately for each packing fraction considered. Figure \ref{figsoft} shows a log-log plot (based on the data published in Fig. 1 of Ref. \cite{Vitelli10} ) of 
$d (\omega) \!\!\!\!\! \quad \omega^4$ versus packing fraction $\Delta \phi$ for the lowest lying group of $12$ modes at $\Delta \phi=0.3$ , $0.1$ , $0.05$ , $0.02$  and $0.01$ (circles) and the second group of 24 modes at $\Delta \phi=0.1$ , $0.05$ and $0.02$ (squares). The continuous line was drawn with slope $11/4$, as predicted in Eq. (\ref{eq:def7}). This scaling appears to be consistent with the numerical results presented in Fig. \ref{figsoft}, thus providing indirect evidence for the length scale $\ell ^{*}$. Larger system sizes are needed to further test the validity of our findings over a greater range of $\Delta \phi$. The main requirement is that the packing fraction is large enough to ensure that the plane wave frequencies are well below $\omega^{*} \sim (\phi - \phi_c)^{\frac{1}{2}}$, but still sufficiently close to $\phi_{c}$ to be in the critical regime.   

\section{\label{test2} Onset of the strong scattering regime}

Above $\omega^{*}$, the scaling analysis of Eq. (\ref{eq:def7}) does not apply because the diffusivity plateaus to a minimal value $d_0$ independent of packing fraction \cite{Xu09,Vitelli10}. This plateau starts at $\omega^{*}$ and ceases to exist only at the very end of the spectrum where localization sets in and the diffusivity becomes vanishingly small.

We will now provide a brief summary of a scaling analysis detailed in Ref. \cite{Vitelli10} for the packing fraction dependence of $\omega^{*}$ that marks the onset of the plateaus in the diffusivity and density of states. Since $d_{0}$ is weakly dependent on $\Delta \phi$ the mean free path at the onset of the diffusivity plateau is given by \cite{Vitelli10}
\begin{equation}
\ell_{d} \sim \frac{d_{0}}{v_t} \sim \frac{1}{{\Delta \phi}^{1/4}} \!\!\! \quad ,
\label{eq:d2}
\end{equation} 
where we used Eq. (\ref{eq:def6}) and Eq. (\ref{eq:scaling2bis}). The diverging length scale $\ell_d$ that controls the onset of the diffusive modes
above $\omega^{*}$ is distinct and larger than $\ell^{*}$, for small $\Delta \phi$.

Although $\ell_{d}$ can be arbitrarily large compared to the particle size, it is comparable to the corresponding wavelength $\lambda^{*}$ at $\omega^{*}$. As a result, energy transport is severely hindered above $\omega^{*}$.
The validity of this heuristic criterion for the onset of strong scattering, $\ell_{d} \sim \lambda^{*}$, has been corroborated by numerical tests in Ref. \cite{Xu09,Vitelli10}.
This allows one to readily convert the wavelength (of order $\ell _{d}$) to the corresponding frequency $\omega^{*}$ by using the phonon dispersion relation for transverse phonons. The result reads
\begin{equation}
\omega^{*} \sim  {v_t}^{2} \sim \Delta \phi ^{\frac{1}{2}} \!\!\! \quad .
\label{eq:do3b}
\end{equation} 

A natural question that arises in this context is how the low frequency regime of diverging diffusivity "meets" the plateau in $d(\omega)$ expected above $\omega^{*}$. The plot of the diffusivity obtained from the numerical evaluation of the Kubo formula in Ref. \cite{Xu09,Vitelli10} lacks the 
spectral resolution necessary to probe in detail the crossover region. Recent mean field calculations carried out on lattice models, that mimic several of the distinctive properties of soft spheres packings, predict that the pre-factor of Rayleigh's law scales as ${\Delta \phi}^{7/4}$ for $\omega \ll \omega^{*}$ \cite{Mattnew}.  This result differs from the ${\Delta \phi} ^{11/4}$ scaling obtained in Eq. (\ref{eq:def7}). This discrepancy may originate from the fact that the mean field model of Ref. \cite{Mattnew} does not capture the length scale $\ell ^{*}$ which is the crucial ingredient that enters our scaling argument leading to Eq. (\ref{eq:def7}). 

In the mean field calculations, the Rayleigh scattering regime extends almost all the way to frequencies of order $\omega^{*}$ so that $d_{MF}(\omega^{*})$, the mean field diffusivity at $\omega^{*}$, is given by   
\begin{equation}
d_{MF}(\omega^{*}) \sim \frac{{\Delta \phi}^{7/4}}{{\omega^{*}}^{4}} \sim \frac{1}{\Delta \phi ^{\frac{1}{4}}} \!\!\! \quad .
\label{eq:comp}
\end{equation} 
This implies that, as the frequency is lowered through $\omega^{*}$, the diffusivity exhibits an upward jump of magnitude ${\Delta \phi ^{-\frac{1}{4}}}$ which becomes arbitrarily large as $\phi$ approaches $\phi_c$. 
By contrast, $d_{EX}(\omega^{*})$, the value of the diffusivity extrapolated to $\omega^{*}$ according to Eq. (\ref{eq:def7}) reads 
\begin{equation}
d_{EX}(\omega^{*}) \sim \frac{{\Delta \phi}^{11/4}}{{\omega^{*}}^{4}} \sim \Delta \phi ^{\frac{3}{4}} \!\!\! \quad .
\label{eq:comp}
\end{equation} 
Note that there is no reason why the Rayleigh regime ought to extend all the way to $\omega^{*}$. As this critical frequency is approached, the scattering becomes stronger and the assumptions under which Eq. (\ref{eq:def7}) was derived no longer hold. Nonetheless, if such an extrapolation is made, the intercept of the low $\omega$ Rayleigh branch of the diffusivity will fall below the plateau value $d_{0}$, eventually becoming vanishingly small as $\phi$ approaches $\phi_c$. This scenario would imply a non-monotonic $\omega$ dependence of the diffusivity analogous to the behavior of the participation ratio \cite{Xu2010}.
  
   
\section{\label{test} Conclusions}

In this article, we have shown that the elastic attenuation coefficient $\alpha(\omega)$ of long-wavelength shear sound waves exhibits critical scaling near point J according to the power law $\alpha(\omega)\sim \omega^4 (\phi - \phi_c)^{-5/2}$ and diverges as $\phi$ approaches  $\phi_c$. A simple scaling argument has been proposed to obtain the packing fraction dependence of the attenuation coefficient which relies on two crucial physical ingredients: the scaling of the shear modulus $G$ that vanishes at point J and the existence of a characteristic length scale $\ell ^{*}$ that characterizes the spatial fluctuations in the structure and diverges at $\phi=\phi_c$. The critical exponent suggested in this work is consistent with numerical evaluations of the Kubo formula for the energy diffusivity at low $\omega$. The favorable comparison between our scaling argument and numerics supports  the simple view that a jammed solid can be regarded as a mosaic composed of domains of characteristic size given by $\ell^{*} \sim (\phi - \phi_c)^{-\frac{1}{2}}$. We hope that the effects discussed in this work will stimulate further experimental investigations of the dependence of sound attenuation in emulsions and granular media under loading.  
  
\begin{acknowledgments}
I thank N. Xu, M. Wyart, A. J. Liu and S. R. Nagel for helpful discussions. I am grateful  to B. Tighe, M. van Hecke, Y. Shokef and W. T. M. Irvine for their comments. This work was initiated during my stay at the Weizmann Institute
under the auspices of the Feinberg Foundation Visiting Faculty Program Fellowship which I gratefully acknowledge.  

\end{acknowledgments}

\end{document}